\documentclass[a4paper,prl,12pt]{revtex4}
\usepackage{epsfig}
\begin{document}
\title{Dynamics of Crystal Formation in the Greenland NorthGRIP Ice Core}

\author{Joachim Mathiesen, Jesper Ferkinghoff-Borg,
  Mogens H. Jensen, Mogens Levinsen, and Poul Olesen\\}
 \affiliation{The Niels Bohr Institute, Blegdamsvej 17, DK-2100 Copenhagen, Denmark.}
\author{Dorthe Dahl-Jensen and Anders Svensson}%
\affiliation{Department of Geophysics, The Niels Bohr Institute for Astronomy, Geophysics, and Physics, Juliane Maries Vej 30, DK-2100 Copenhagen, Denmark.}%
\begin{abstract}
The North Greenland Ice Core Project (NorthGRIP) provides
paleoclimatic information back to at about 120 kyr before present
({Dahl-Jensen and others}, 2002). Each year, precipitation on the ice sheet
covers it with a new layer of snow, which gradually transforms into
ice crystals as the layer sinks into the ice sheet. The size
distribution of ice crystals has been measured at selected depths in
the upper 880 m of the NorthGRIP ice core ({Svensson and others}, 2003b),
which covers a time span of 5300 years. The distributions change with
time toward a universal curve, indicating a common underlying physical
process in the formation of crystals. We identify this process as an
interplay between fragmentation of the crystals and diffusion of their
grain boundaries. The process is described by a two-parameter
differential equation to which we obtain the exact solution. The
solution is in excellent agreement with the experimentally observed
distributions.
\end{abstract}
\maketitle
\pagebreak
The NorthGRIP drilling location (75.10 N, 42.32 W) is situated on an
ice ridge with an ice thickness of 3085 m. The mean annual temperature 
is -32 deg. Celsius and the annual accumulation in the area is on the average 19.5 cm ice equivalent. In the upper 80 m
of the ice sheet, the firn, the snow gradually compacts to
a close packing of ice crystals of typical sizes 1 to 5 mm. We apply
crystal size distributions obtained from fifteen vertical thin
sections of ice evenly distributed in the depth interval 115 - 880
m ({Svensson and others}, 2003b). The thin sections have dimensions of
20 cm $\times$ 10 cm (height $\times$ width) and a thickness of
0.4$\pm$0.1 mm. 

\begin{figure}[!htbp]
  \begin{center}
   \epsfig{width=7cm,file=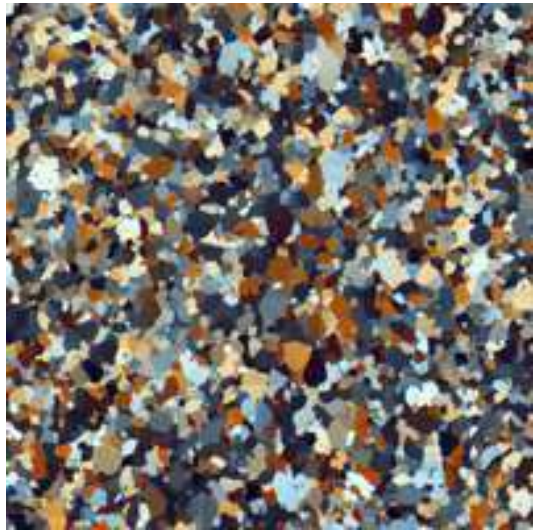}
    \caption{An image of a 10 x 10 cm$^2$ vertical thin section of ice from the depth 115 m. 
The section is viewed between two crossed linear polarizers
 and the different colors represent almost 3000 individual crystals
 with various orientations of the crystal optical c-axes.}
    \label{fig:1}
  \end{center}
\end{figure}
Digital images of ice thin sections placed between crossed linear
polarizers have been used to map the dimensions of individual ice
crystals in the sample (Fig.$\!$ \ref{fig:1}). The ages of the
considered samples are all less than 5300 years ({Johnsen and others},
2001), and the temperature of the ice in this period can be assumed
constant ({Dahl-Jensen and others}, 1998).

\begin{figure}[!htbp]
  \begin{center}
    \epsfig{width=7cm,file=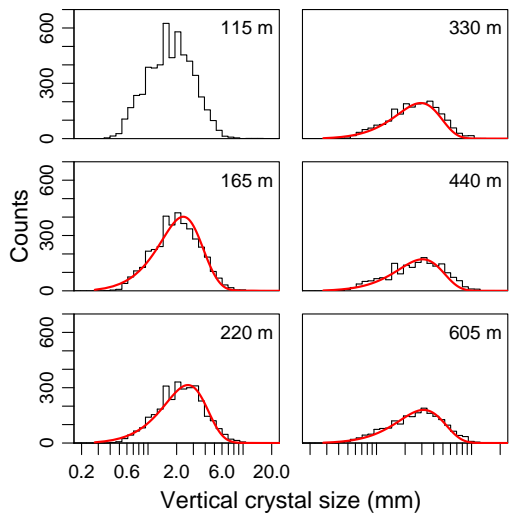}
  \end{center}
  \caption{Distributions of ice crystal sizes at
    depths 115m, 165m, 220m, 330m, 440m and 605m. The crystal size is
    defined as the maximum vertical extent of the individual
    crystals.  The black lines are the measured histograms and the smooth 
lines are the temporal evolution predicted by eq. (\ref{diffeq}) starting from
    the initial distribution at 115m. The total counts of ice crystals
decreases with depth (due to the overall increase of sizes) until the steady state
is reached.}
  \label{fig:2}
\end{figure}
Fig.$\!$ 2 shows size distributions of ice crystals at selected depths
down to 880 m (5300 years before present (B.P.)). The crystal size is
defined as the vertical extent of a crystal, which is
estimated as the height of the minimal and vertical aligned
rectangle, which encloses the individual crystal. Using the
horizontally measured sizes of the ice crystals from the thin
sections, one obtains similar distributions.  
The thin sections used in determining the ice size
distributions have typical thicknesses of 0.4$\pm 0.1$ mm, and
ice crystals with all length scales smaller than this
thickness will consequently be either smoothed away when cutting the ice or be
invisible when viewed in the linear polarizers. These small crystals
will be seen as part of the larger crystals and not as individual
crystals. This corresponds to an average increase in the
sizes of the large crystals, i.e. the lengths we measure $x$ are equal
to the real length $\hat x$ plus some noise $\epsilon$ of positive
mean, $x=\hat x+\epsilon$. We have assumed that this noise is
sufficiently peaked around half the thickness of the thin section, such
that we simply put $\epsilon=0.2$mm. This value of $\epsilon$ has been
used in the comparisons between theoretical and experimental values.

Each distribution exhibits a
pronounced peak, indicating a typical crystal size at each depth,
followed by an exponentially decaying tail of relatively large
crystals. The mean size becomes
larger with depth and thus time until it reaches a limiting crystal
size ({Thorsteinsson and others}, 1997; {Li and others}, 1998).  The
most important physical process behind such growth is  diffusion
of grain boundaries between the ice crystals. Inhomogeneities on the
crystal boundaries, characterized by small radii of curvature,  tend
to be smoothened out as time progresses. This causes the inhomogeneities to be incorporated
into the larger crystals leading to an overall growth of the mean crystal size
({Alley and others}, 1986; {Paterson}, 1994). This approximative
description leads to the so-called Normal Grain Growth law that
predicts the following temporal dependence of the mean crystal size
$x$, 
\begin{equation}
  \label{eq:1}
 \langle x^2 \rangle(t) = x_0^2 + kt, 
\end{equation}
 where $x_0$ is the initial mean crystal size and $k$ is the crystal
 growth rate. This approach was successfully applied for the GRIP ice
 core by {Thorsteinsson and others}, 1997, who fitted their
 experimental observations well back to 2500 B.P. By definition,
 however, the diffusion does not cause the crystal size to reach a
 limiting size, which is in disagreement with the actual observations
 that clearly indicates a saturation in $\langle x^2 \rangle$ after
 2500 B.P. ({Alley and Woods}, 1996; {Gow and others}, 1997; {Thorsteinsson and others}, 1997).

The important missing part of the physical mechanism
in this model is the fact that all crystals, and in particular the
large ones, are subjected to fragmentation processes or - as often
referred to in ice physics - polygonization or rotation
recrystallization ({Alley and others}, 1995; {Thorsteinsson et
  al.}, 1997). We here include this process in the description which
then leads to a physical model with a balance between the diffusion of
grain boundaries and the fragmentation of crystals causing the mean
crystal size to reach a steady state after 2500 B.P.

Another process that has been proposed as important for the production
of small crystals is nucleation, or the formation of new grains, which
may have c-axis orientations that differ from the dominating vertical
orientation of the surrounding crystals. Measurements of the NorthGRIP
fabric (c-axis orientations) suggest, however, that nucleation does
not seem to play an important role in the Holocene ({Wang and others},
2002; {Svensson and others}, 2003b), and nucleation is not included in
the model.

Our proposed model is formulated as
a rate equation in the quantity $N(x,t)$, which is the density of
ice crystals of length $x$ at time $t$ measured B.P. In this study $x$ is chosen as the vertical extent of the crystals. At a given time, $N(x,t)$ can
be increased or decreased by diffusion with a diffusion constant $D$. It can
receive fragments of size $x$ from fragmentations of larger crystals and 
it can decrease by its own fragmentation. The fragmentation is
defined as a rate $f$ in length and time, i.e. for a given time step
$dt$ the average number of fragmentation events over a length
$L$ is $fLdt$. 

The fragmentation rate $f$, and the diffusion constant $D$, will
depend on temperature, non-hydrostatic stresses in the ice and 
moreover be sensitive to the impurity content of the ice ({Alley and others},1986). 

These diffusion and fragmentation processes lead
to an integral-differential equation on the form
\begin{equation}
\label{diffeq}
  \frac{\partial N(x,t)}{\partial t}=D\frac{\partial^ 2 N(x,t)
  }{\partial x^ 2}-fxN(x,t)+2f\int_x^\infty N(x',t)dx'
\end{equation}

Here, the first term on the right hand side is a diffusion term that
corresponds to the grain boundary diffusion of the Normal Grain Growth
law, i.e. if we only include this term in the model we reproduce for
large times the parabolic behavior in (\ref{eq:1}). Note that the equation describes
the dynamics of the full assembly of crystals and that the mean square of
the crystal sizes therefore follows from
\begin{equation}
  \label{eq:2}
  \langle x^2\rangle(t)=\int_0^\infty x^2N(x,t)dx\left/\left(\int_0^\infty
  N(x,t) dx\right)\right. .
\end{equation}
If we only include the diffusion term in
eq. (\ref{diffeq}), we get for large times that $\langle
x^2\rangle(t)\sim 4Dt$. The behavior at large times differs from
normal diffusion by a factor of two because we use the absorbing boundary
condition that for all times $N(0,t)=0$. 

The last term in (\ref{diffeq}) is the contribution from
fragmentation of larger crystals into crystals of size $x$. Combining
the two ways a crystal of size
$x'>x$ can produce a fragment of size $x$ with the assumption that there is a
uniform probability, $1/x'$ for where the crystal break,
we get
\begin{equation}
  \label{eq:3}
fx'N(x')\cdot\frac 2 {x'},  
\end{equation}
where $fx'N(x')$ is the number of crystals of size $x'$ that fragments
per time and $2/x'$ is the probability for generating a fragment of size
$x$. Note that the assumption of a uniform probability already was
made in the definition of the rate $f$. If we integrate (\ref{eq:3}) over all crystals larger than $x$ we
achieve the last term in (\ref{diffeq}).

\begin{figure}[htbp]
  \begin{center}
    \epsfig{height=7cm,file=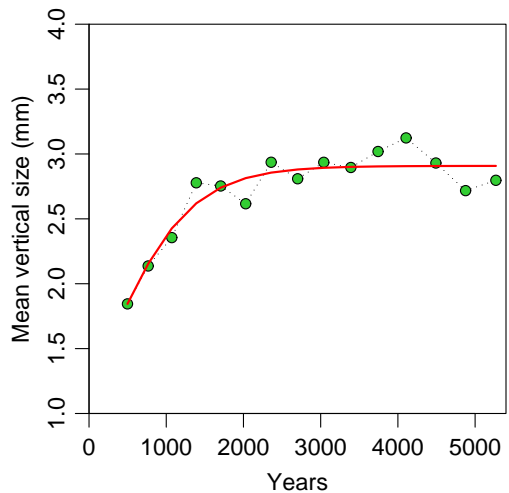}
  \end{center}
  \caption{The mean vertical size of the ice crystals shown
    versus their age in years B.P. The smooth line shows the best fit predicted
    from our dynamical description of ice crystal growth. From the fit
    we read off the diffusion constant, $D\approx1.4\cdot
    10^{-3}$mm$^2\cdot$yr$^{-1}$, and fragmentation rate,
    $f\approx5.2\cdot 10^{-4}$yr$^{-1}\cdot$mm$^{-1}$. The time scale is taken from ref. ({Johnsen and others}, 2001).}
  \label{fig:3}
\end{figure}

Dividing through by $f$ in eq. (\ref{diffeq}), we obtain an expression
on the right hand side which depends on one parameter $a=D/f$
only. The integral-differential equation
has analytic solutions, $B(\frac{x+\lambda}{a^{1/3}})$, which are explicitly given in terms of the Bessel
function $K_{\frac{1}{3}}$ and eigenvalues $\lambda$,
\begin{equation}
B\left(\frac{x+\lambda}{a^{1/3}}\right)= \frac{x^ {3/2}}{\sqrt{3}}K_{\frac{1}{3}}\left(\frac{2}{3}\left(\frac{x+\lambda}{a^{1/3}}\right)^{3/2}\right).
\end{equation}
The function $K_{\frac{1}{3}}(x)$ can be written as a sum of the more
frequently used Bessel functions $I(x)$, $K_{\frac{1}{3}}(x)= \frac{\pi}{
\sqrt 3}(I_{-\frac{1}{3}}(x)-I_{\frac{1}{3}}(x))$.
 The boundary condition, $N(0,t)=0$ for all times $t$, implies that only a discrete set of
non-positive values for $\lambda$ is allowed. They are found by
solving the equation $ B\left(\frac\lambda{a^{1/3}}\right)=0.$ The
largest eigenvalues are $\lambda_0/a^{1/3}=0,~\lambda_1/a^{1/3}=-2.338\ldots,~\lambda_2/a^{1/3}=-4.088\ldots,\ldots.$
 The general solution can be written as a linear combination of the
eigenfunctions, 
\begin{equation}
N(x,t)=\sum_{n=0}^\infty c_nB\left(\frac{x+\lambda_n} {a^{1/3}}\right) e^{\lambda_nf(t-t_0)}
\end{equation}
where the $c_n$'s can be computed from the distribution at time
$t=t_0$  (for further details see the work of ({Ferkinghoff-Borg
  and others}, 2003)). The fact that we have no positive eigenvalues guarantees 
the existence of a steady state solution, which is defined by the
only surviving term (corresponding to $\lambda_0=0$) for high ages,
\begin{equation}
  \label{eq:4}
N(x,t)\sim B(x/a^{1/3}) ~~~~~\mbox{for\quad $t\rightarrow \infty$.}  
\end{equation}
The characteristic time, $\tau$, of
 the exponential growth towards steady state is given by the second
 largest eigenvalue, $\lambda_1$, i.e.
$$
\tau=-1/(\lambda_1 f)\approx(2.338\cdot f\cdot a^{1/3})^{-1}.
$$  
When the dynamics has reached steady state, and the mean crystal size
has saturated, the distribution is described by the single parameter,
$a^{1/3}$, which defines the characteristic length scale of the
system. In particular, one finds that the mean vertical size in the steady state is
$$
\langle x\rangle_\infty=3^{2/3}\frac{\Gamma(4/3)}{\Gamma(3/2)}a^{1/3}.
$$ 
In Fig.$\!$ \ref{fig:3} we show the mean vertical size of the ice
crystals, $\langle x\rangle(t)$ as function of time. The dots are the
experimental values and  the solid line is an exponential fit
corresponding to the two leading terms in the solution, 
\begin{equation}
\frac{\langle x\rangle_\infty}{1+\left(\frac{\langle x\rangle_\infty}{\langle x \rangle_0}-1\right)e^{-(t-t_0)/\tau}},
\end{equation}
where $\langle x\rangle_0$ is the observed average length at time
$t_0=500$ years. 
From the figure, we estimate the characteristic time, $\tau=
600\pm 70$ years and the average length in the steady state
$\langle x\rangle_\infty=2.9\pm 0.1$ mm. The two parameters correspond
to an effective fragmentation rate and a diffusion constant of
respectively $f= (5.2\pm 0.6)\cdot 10^{-4}$ mm$^{-1}\cdot$yr$^{-1}$ and
$D=(1.4\pm 0.2)\cdot 10^{-3}$ mm$^2$yr$^{-1}$. Following eq. (3) we
obtain the Normal Grain Growth crystal growth rate $k =4 D = (5.5\pm 0.8)\cdot
10^{-3}$mm$^2$yr$^{-1}$, which corresponds very well with the known
values for GRIP, $k = 5.6 \cdot10^{-3}$mm$^2$yr$^{-1}$ ({Thorsteinsson and others}, 1997),
and for NorthGRIP, $k = 5.8 \cdot10^{-3}$mm$^2$yr$^{-1}$ ({Svensson and others}, 2003b). In this work
we are using the maximum vertical extent of the crystals, so that no
correction value for the sectioning effect is needed.

Using these estimates, we can predict the time evolution of crystal sizes from any initial
distribution. The solid lines in Fig.$\!$ \ref{fig:2} show the
time evolution of the distribution observed at time $t_0=500$ years (115 m depth), in good agreement with the experimental results. 

When comparing the model to the vertical crystal sizes we should in principle take into account the vertical compression of crystals due to the ice flow. Had we chosen the horizontal crystal sizes for comparison, we should have taken into account the corresponding elongation. However, observations show that the crystal flattening, which is defined as crystal width divided by crystal height, is less than $15\%$ for the applied samples ({Svensson and others}, 2003b), so in order to keep things reasonable simple, we simply ignore this effect.

\begin{figure}[htbp]
  \begin{center}
    \epsfig{width=7cm,file=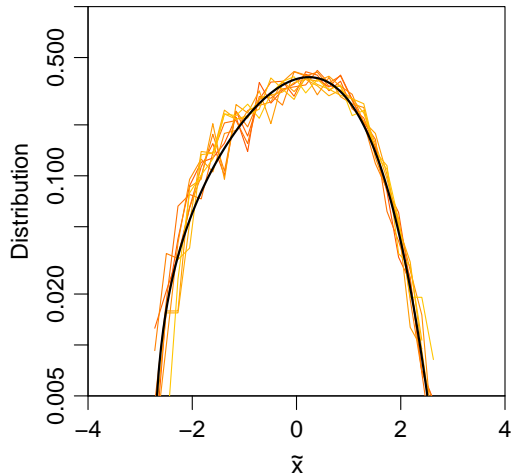}
  \end{center}
  \caption{The figure shows a ``data collapse''  of the size
    distributions as a consequence of a rescaling, $\tilde x = (\log
    x-\langle \log x \rangle)/\sigma(\log(x))$, i.e. the shown
    distributions have zero mean and unit standard deviation. The
    lines correspond to the eight data points in Fig. \ref{fig:3} of the oldest samples ($t>2500$ years) and the
    black line on top is the steady-state solution of eq.(1). We use
    the rescaling $\tilde x$ in order to improve the resolution around the smallest crystal
    sizes and note that the steady-state solution is transformed accordingly.}
  \label{fig:4}
\end{figure}

During an annual cycle, the impurity content and the average
crystal size show important variations with the highest dust load and
the smallest crystals appearing during spring ({Svensson and others},
2003a). Because the size of the applied samples (20 cm depth) is close
to the annual accumulation at NorthGRIP (19.5 cm ice equivalent), the
seasonal variability will to a first order be averaged out in the experimental
data. Still, some inter-annual variations in the impurity content of
the ice are observed ({Svensson and others}, 2003a). These
fluctuations may explain the small discrepancies between model and
observations in Figs. \ref{fig:2} and \ref{fig:3}, since the values of
$D$ and $f$ are determined by the average concentration. We can take
these fluctuations into account for the older samples, which have
almost reached a steady state, by rescaling the crystal size
distributions. In Fig.$\!$ \ref{fig:4} we demonstrate that
this rescaling results in a universal curve for all the size
distributions, described by the steady state Bessel function solution,
which is significantly different from the widely used log-normal
distribution. The data-collapse in the figure, i.e. the fact that all
the rescaled distributions have the same shape, clearly supports the
one-parameter nature of the steady state solution (\ref{eq:4}). 

By formulating eq. (\ref{diffeq}) we thus provide a comprehensive description
of the dynamical processes of ice crystals. Our model improves the
Normal Grain Growth law based on diffusive growth alone and explains
the reach of a steady state for the mean crystal size by means of
fundamental physical processes. The very good agreement between data
and model suggest that polygonization or fragmentation is the
dominating process in production of small crystals rather than
nucleation of new crystals, which is a process not included in the
model. The suggested interplay between the diffusion  and the
fragmentation in the crystal dynamics is believed to 
be a central ingredient in many other systems in nature (like
in geological and perhaps biological processes, see ({Ferkinghoff-Borg
  and others}, 2003)) and could provide 
a useful tool to predict the distribution curves of fragmented pieces.

\end{document}